# Hand tracking for immersive virtual reality: opportunities and challenges

**Gavin Buckingham**
Department of Sport and Health Scineces
University of Exeter
Exeter, UK

**Abstract**

Hand tracking has become an integral feature of recent generations of immersive virtual reality head-mounted displays. With the widespread adoption of this feature, hardware engineers and software developers are faced with an exciting array of opportunities and a number of challenges, mostly in relation to the human user. In this article, I outline what I see as the main possibilities for hand tracking to add value to immersive virtual reality as well as some of the potential challenges in the context of the psychology and neuroscience of the human user. It is hoped that this paper serves as a roadmap for the development of best practices in the field for the development of subsequent generations of hand tracking and virtual reality technologies.

Immersive virtual reality (iVR) systems have recently seen a huge growth due to reductions in hardware costs and a wealth of software use cases. In early consumer models of the Oculus Rift Head-Mounted Display (HMD), interactions with the environment (a key hallmark of iVR) were usually performed with hand-held controllers. Hands were visualized in games and applications (infrequently) in a limited array of poses based on finger position, assumed from contact with triggers and buttons on these controllers. Although the ability to visualize the positions of individual digits was possible with external motion tracking and/or 'dataglove' peripherals which measured finger joint angles and rotations, these technologies were prohibitively expensive and were unreliable without careful calibration. A step change in hand tracking occurred almost accidentally with the Leap Motion Tracker, a small encapsulated infra-red emitter and camera optical geared toward the (apparently futile) goal of having people interacting with desktop machines by gesturing at the screen. This device was very small, required no external power source, and was able to track the movements of individual digits in three dimensions using a stereo camera system with reasonable precision (Guna et al., 2014). Significant improvements in software, presumably through a clever use of inverse kinematics, along with a free software-development kit and a strong user base in the Unity and Unreal Game Engine communities led to a proliferation of accessible hand tracking addons and experiences tailor-made for iVR. The developers of Leap Motion Tracker eventually capitalized on this device's unforeseen potential for iVR, pivoting the device exclusively to iVR use cases. Since then, hand tracking has become embedded into the hardware of recent generations of iVR HMDs (e.g., the first and second iterations of the Oculus Quest) through so-called 'inside out' tracking, and looks set to continue to evolve with emerging technologies such as wrist-worn electromyography ("Inside Facebook Reality Labs," 2021). This paper will briefly outline the main use-cases of hand tracking in VR for the purposes of visualization, and then discuss in some detail the outstanding issues and challenges which developers need to keep in mind when developing such experiences.

Opportunities - Why hand tracking?

Our hands, with the dexterity afforded by our opposable thumbs, are one of the canonical features which separates us from non-human primates. We use our hands to gesture, feel, and interact with our environment almost every minute of our waking lives. When we are prevented from, or limited in, using our hands, we are profoundly impaired, with a range of once-mundane tasks becoming frustratingly awkward. In the context of iVR, the primary value of effective hand tracking is for visualization of the hands within a virtual environment, which can itself be divided into three (not unrelated) categories: immersion, communication, and interaction.

For the first, immersion, being able to see your hands animated in real time adds a significant wow factor to your experience. Research has shown that we have an almost preternatural sense of our hands' positions and shape when they are obscured (CITATION), and when our hands are removed from our visual worlds it is a start reminder of our disembodiment. Indeed, we spend the majority of our time during various mundane tasks foveating our hands (CITATION), so removing them from the visual scene presumably has a range of consequences for our visuomotor behaviour.

Which leads to the next point: interaction. A key point of virtual reality is the ability to interact with the computer-generated environment. This can be simply through moving the head to experience the wide visual world, but modern VR experiences usually involve some form of manual interaction, from opening doors to wielding weapons. Accurate tracking of the hands allows for the embodiment of these experiences, adding not only to the user's immersion, but presumably also the accuracy of their movements, which seems particularly key in the context of training.

The final point to discuss is that of manual gesticulation – the use of one's hands to emphasize words and punctuate sentences through a series of gestures. 'Gestures' in the context of HCI has come to mean the swipes and pinching motions uses to perform commands. However, the involuntary movements of hands during natural communication appear to play a significant role not just for the listener, but also the communicator (the latter to such an extent that conversations between two congenitally blind individuals contain as many gestures as conversations between sighted individuals). Indeed, recent research has shown that individuals are impaired in recognizing a number of key emotions in the images of bodies which have the hands removed (Ross and Flack, 2020), highlighting how important hand form information is in communicative experiences. The value of manual gestures for communication in virtual environments is compounded given that veridical real-time face tracking and visualisation is technically very difficult due to the extremely high temporal and spatial resolution required to detect and track microexpressions. Furthermore, computer-generated faces are particularly prone to large uncanny-valley like effects whereby faces which fall just short of being realistic elicit a strong sense of unease (MacDorman et al., 2009; McDonnell and Breidt, 2010). Significant recent strides have been made in tracking and rendering photorealistic faces (Schwartz et al., 2020), but the hardware costs are likely to be prohibitive for the current generation of consumer-based VR technologies. Tracking and rendering of the hands, with their large and expressive kinematics, should thus be strong a focus for communicative avatars in the short term.

Challenge 1 - Object interaction

Our hands are one of our main ways to effect change in the environment around us. Thus, one of the main reasons to visualise hands in VR is to facilitate and encourage interactions with the virtual environment. From opening doors to wielding weapons, computer-generated hands are an integral part of many game experiences across many platforms. As outlined above, these manual interactions are typically generated by reverse-engineering interactions with a held controller. For example, on the Oculus Quest 2 controller, if the buttons underneath the index and middle fingers are lightly depressed, the hand appears to close slightly; if the buttons are fully depressed, the hand closes into a fist. Not only does this method of interacting with the world feel quite engaging, it elicits a greater sense of ownership over the seen hand than a visualization of the held controller itself (Lavoie and Chapman, 2021). But despite the compelling nature of this experience, hand tracking offers the promise of a real-time veridical representation of the hands' true actions, requiring no mapping of physical to seen actions and untethered from any extraneous hardware.

But it is far from clear whether this promise is realised in a beneficial way for usability. Anecdotally, interacting with virtual objects using hand tracking feels imprecise and difficult to use, which is supported by recent findings showing that during a block moving task hands tracked with a Leap Motion tracker score lower on the System Usability Scale than hands tracked with a hand-held controller (Masurovsky et al., 2020). Furthermore, subjective Likert ratings on a number of descriptive metrics suggested that the controller-free interaction felt significantly less comfortable and less precise than the controller-based interactions. Even more worryingly, this same article noted that participants performed worse on a number of performance metrics when their hands were tracked with the Leap than with the controller.

It is likely that the main reason that controller-free hand tracking is problematic during object interaction is the lack of tactile and haptic cues in this context. Tactile cues are a key part to successful manual actions, and their removal impairs the accuracy of manual localization (Rao and Gordon, 2001), alters grasping kinematics (Furmanek et al., 2019; Mangalam et al., 2021; Ozana et al., 2020; Whitwell et al., 2015), and affects the normal application of fingertip forces (Buckingham et al., 2016). While controller-based interactions with virtual objects do not deliver the same tactile and haptic sensations experienced when interacting with objects in the physical environment, the vibro-tactile pulses and the mass of the controllers do seem to aid in scaffolding a compelling percept of touching something. A range of solutions to replace tactile feedback in the context of VR have been developed in recent years, from glove-like devices which provide tactile feedback and force feedback to the digits (Carlton, 2021) to stimuli which precisely deform the fingertips to create a sensation of the mechanics of interaction (Schorr and Okamura, 2017) to devices which deliver contactless ultrasonic pulses aimed at the hands to simulate tactile cues (Rakkolainen et al., 2019). It is currently unclear which of these solutions (or one hitherto unforeseen ) will solve this challenging issue, but it clearly is the foremost challenge in the broad uptake immersive virtual reality.

Challenge 2 - Tracking location

With 'inside-out' cameras in current consumer models (e.g., the Oculus Quest 2), hand tracking is at its most reliable when the hands are roughly in front of the face, presumably to maximise the overlap of the fields of view of the individual cameras which track the hands. In these headsets, the orientation of these cameras is fixed, presumably due to the assumption that participants will be looking at what they are doing in VR. This assumption is probably appropriate for discrete game-style 'events' – it is well-established that individuals foveate the hands and the action endpoint during goal-directed tasks (Desmurget et al., 1998; Johansson et al., 2001; Lavoie et al., 2018). In more natural sequences of tasks (e.g., preparing food), however, the hands are likely to spend significant proportion of time in the lower visual field due to their physical location below the head. This asymmetry in the common locations of the hand during many tasks was discussed in the context of a lower visual field specialization for manual action by Previc (1990) and has received support parallels from a range of studies showing that humans are more efficient utilizing visual feedback to guide effective reaching toward targets in their lower visual field than their upper visual field (Danckert and Goodale, 2001; Khan and Lawrence, 2005; Krigolson and Heath, 2006). This behavioural work is supported by evidence from the visual system for a lower visual field speciality for factors related to action (Schmidtmann et al., 2015; Zhou et al., 2017), as well as neuroimaging evidence that grasping objects in the lower visual field preferentially activates a network of dorsal brain regions specialised for planning and controlling visually-guided actions (Rossit et al., 2013). As the range of tasks undertaken in VR widens to include more natural everyday experiences where the hands might be engaged in tasks in the lower visual fields of, limitations of tracking and visualization in this region of space will likely become more apparent. Indeed, this issue is not only one of tracking, but hardware field of view. Currently the main focus on field of view is concerned with increasing the lateral extent, with little consideration given to the fact that the 'letterbox' shape of most VR HMDs reduce the vertical field of view in the lower visual field by more than 10% compared to that which the eye affords in the physical environment (okreylos, 2019, 2016). Together, these issues of tracking limitations and physical occlusion are likely to result in unnatural head movements in manual tasks to ensure the hands are kept in view which could limit the transfer of training from virtual to physical environments, or significant impacts on immersion as the hands disappear from peripheral view at an unexpected or inconsistent point.

Challenge 3 - Uncanny phenomenon and embodiment

The uncanny phenomenon (sometimes referred to as the uncanny valley) refers to the lack of affinity yielding feelings of unease or disgust when looking at, or interacting with, something artificial which falls just short of appearing natural (Mori, 1970; Wang et al., 2015). The cause of this effect is still undetermined, but recent studies have suggested that this effect might be driven by mismatches between the apparently-biological appearance of the offending stimuli and non-biological kinematics and/or inappropriate features such as temperature and surface textures. (Kätsyri et al., 2015; Saygin et al., 2012). The main triggers for uncanny valley seem to be in the realms of computer-generated avatars (MacDorman et al., 2009; McDonnell and Breidt, 2010) and interactive humanoid robots (Destephe et al., 2015; Strait et al., 2017) and, as such, much of research into this topic has focussed on faces. Recent studies have suggested that this effect is amplified when experienced through an HMD (Hepperle et al., 2020), highlighting the importance of this factor in the context of tracked VR experiences.

Little work has, by contrast, has examined such responses toward hands. In the context of prosthetic hands, Poliakoff and colleagues (2018, 2013) demonstrated that images of life-like prosthetic hands were rated as more eerie than anatomical or robotic hands in equivalent poses. This effect appears to be eliminated in some groups with extensive experience (e.g., in observers who themselves have a limb absence), but is still strongly experienced by prosthetists and non-amputees trained to use a prosthetic hand simulator (Buckingham et al.,

2019). Given the strong possibility of inducing a presence-hindering effect if virtual hands are sufficiently disconcerting (Brenton et al., 2005), it seems prudent to recommend outline or cartoon hands as the norm for even strongly-embodied VR experiences. This suggestion is particularly important for 'untethered' HMDs, due to the fact that rendering photorealistic images of hands tracked at the high frequencies required to visualize the full range of dextrous actions will require significant computing power. A final point in this regard which also bears mention is that the uncanny valley is not a solely visual experience, but a multisensory one. For example, it has been shown that users subjective experience of an experience rapidly declines when the visual cues in a VR scenario do not match with the degree to haptic feedback (Berger et al., 2018). Furthermore it has recently been shown that when the artificiality of tactile cues and visual cues are mismatched, this can also generate a reduction in feelings of ownership (D'Alonzo et al., 2019). Thus if tactile cues are to become a feature of hand tracking and visualization, care must be taken to avoid features of this so-called 'haptic uncanny valley' (Berger et al., 2018).

Challenge 4 - Inclusivity

Inclusivity is an increasingly important ethical issue in technology, and the development of hand tracking and visualization in iVR throws up a series of unique challenges in this regard. A fundamental part of marker-free hand tracking is to segment the skin from the surrounding background to build, and ultimately visualize, the dynamics of the hand. One potential issue which has not received explicit consideration is that of skin pigmentation. There are a number of recent anecdotal examples (Fussell, 2017) of examples framed around hardware limitations where items from automatic soap dispensers to heart-rate monitors fail to function as effectively for individuals with darker skin tones (which are less reflective) than lighter skin tones (which are more reflective). It is critical that, as iVR is more widely adopted, the cameras which track the hands are able to adequately image all levels of skin pigmentation.

A related issue comes from the software which is used to turn the images captured by the cameras into dynamic models of the hands, using models of possible hand configurations (inverse kinematics). These models, assuming they are built from training sets, are likely to suffer from the same algorithmic bias which has been problematic in face classification research (Buolamwini and Gebru, 2018), with datasets largely derived from Caucasian males yielding startling disparities in levels of misclassification across skin type and gender. This issue becomes one not just of skin pigmentation, but of gender, age, disability, and skin texture and presumably will be exacerbated at these intersections. Any hardware and software which aims to cater for the 'average user' risks leaving hand tracking functionally unavailable to large portions of society. One possible solution to this could be to have users generate their own personalised training sets, akin personalized 'voice profiles' used in some speech recognition software and home assistant devices.

The final issue on this topic relates to the visualization of the hands. Although the current norm for hand visualization is for outline or cartoon-style hands which lack distinguishing features, presumably there will be a drive for the visualization of more realistic-looking hands. As is becoming standard for facial avatars in CG environment, it is important for individuals to be able to develop a model in the virtual environment steps away from the 'default' of an able-bodied Caucasian male or female toward one which accurately represents their bodily characteristics (or, indeed, that of another). This can be jarring – for example it has been shown that the appearance of opposite-gender hands reduces women's experience of presence in virtual environments (Schwind et al., 2017). With hands, this is also likely to be particularly important from an embodiment perspective, with an emerging body of literature suggesting that individuals are less able to embody hands which appear to be from a visibly different skin tone than their own (Farmer et al., 2012; Lira et al., 2017).

Conclusions

In summary, hand tracking is probably here to stay as a cardinal (but probably still optional) feature of immersive virtual reality. The opportunities for facilitating effective and engaging interpersonal communication and more formal presentations in a remote context is particularly exciting for many aspects of our social, teaching, and learning worlds. Being cognisant of the challenges which come with these opportunities is a first step toward developing a clear series of best practices to aid in the development of the next generation of VR hardware and immersive experiences.

Acknowledgements

The author would like to thank João Mineiro for his comments on an earlier draft of this manuscript.